\documentclass[useAMS]{mn2e}

\usepackage{graphicx,subfigure}
\usepackage{pslatex}

\title[Partial Paschen-Back splitting of SiII and SiIII lines]{Partial Paschen-Back splitting of SiII and SiIII lines in magnetic CP stars}
\author[V. Khalack \& J. D. Landstreet]
       {
        V. Khalack$^1$, J. D. Landstreet$^{2,}$$^3$ \\
               $^1$D\'epartement de Physique et d'Astronomie,
               Universit\'e de Moncton, Moncton, N.-B., Canada E1A 3E9\\
               $^2$Department of Physics \& Astronomy, The University of Western Ontario, London, Ontario, N6A 3K7, Canada \\
               $^3$Armagh Observatory, College Hill, Armagh, BT61 9DG, Northern Ireland, UK\\
            }
\date{Accepted ???.
      Received ???;
      in original form ???}

\pagerange{\pageref{firstpage}--\pageref{lastpage}} \pubyear{2010}

\begin{document}

\maketitle

\label{firstpage}

\begin{abstract}
Modelling of the spectra of magnetic A and B main sequence stars is generally done assuming that all the lines are split by the magnetic field according to the Zeeman effect. However, a number of prominent spectral lines are produced by closely spaced doublets or triplets. Such lines should be treated using the theory of the partial Paschen-Back (PPB) effect. Depending on the strength and orientation of magnetic field, the PPB effect can result in the Stokes $I$ and $V$ profiles of a spectral line that differ significantly from those predicted by the Zeeman effect theory. It is important to understand the size and types of errors that are introduced into magnetic spectrum synthesis by treating such lines with the usual Zeeman splitting theory rather than using the correct theoretical treatment of line splitting. To estimate the error introduced by the use of the Zeeman approximation, numerical simulations have been performed for spectral lines of the element silicon, for which a number of important lines are actually in the PPB regime, assuming an oblique rotator model, for various silicon abundances and  $V\sin{i}$ values. A comparative analysis of the Stokes $I$ and $V$ profiles calculated assuming the PPB and Zeeman splitting has been carried out for a number of both strong and weak Si\,{\sc ii} and Si\,{\sc iii} lines.
The analysis indicates that for high precision studies of some spectral lines the PPB approach should be used if the field strength at the magnetic poles is $B_{\rm p}>$ 10 kG. 
%For high signal-to-noise ratio spectra and/or enhanced silicon abundance, the PPB effect can become important even for a field strength of $B_{\rm p}>$ 6 kG.
In the case of the Si\,{\sc ii} line 5041\AA, the difference between the two simulated profiles is caused by a significant contribution from a so called ``ghost" line.
The Stokes $I$ and $V$ profiles of this particular line simulated taking into account PPB splitting provide a significantly better fit to the observed profiles in the spectrum of the magnetic Ap star HD~318107 than the profiles calculated with Zeeman splitting.
Employing the PPB approach, the Si\,{\sc ii} 5041\AA\, line can become a useful tool for abundance mapping and reconstruction of magnetic field configuration, due to its sensitivity to the silicon abundance and to the magnetic field strength.
\end{abstract}


\section{Introduction}

In the presence of a magnetic field, spectral lines become wider due to the splitting of the energy levels
corresponding to the individual transitions. The phenomenon was discovered by Pieter Zeeman \shortcite{Zeeman1897},
after whom it was named the Zeeman effect. Applying the experimentally determined splitting and polarisation characteristics of the Zeeman effect to observed solar spectra, George Ellery Hale \shortcite{Hale1908} was able to demonstrate the presence of a magnetic field in sunspots. Using a similar technique for the analysis of stellar spectra, Horace Babcock \shortcite{Babcock47} discovered Zeeman splitting of absorption lines in circularly polarized spectra of the A2p star 78 Vir and deduced the presence of a 1.5~kG magnetic field in its atmosphere. Nowadays, the Zeeman effect is widely used to estimate the magnetic field strength and configuration in the stellar atmospheres of magnetic stars, some of which also show peculiar chemical abundances (Donati et al. 1997; Mathys \& Hubrig 1997;  Wade et al. 2000a).
% and circumstellar medium (ref.)?

When the Zeeman effect is observed in an atom in a magnetic field that is small enough that the magnetic splitting of the energy levels responsible for the transition is small compared to the fine structure (spin-orbit) splitting of those levels, a wide variety of magnetic splitting patterns is observed. This behaviour is termed (for historical reasons) the "anomalous" Zeeman effect. However, when the field is strong enough that magnetic splitting becomes substantially larger than spin-orbit fine structure splitting, the observed magnetic pattern splitting changes gradually to a simple ("normal") Zeeman triplet. The extreme case of magnetic splitting in this large-field limit was first analysed by Friedrich Paschen and Ernst Back \shortcite{P+B21}, and was later named the Paschen-Back effect.  Fifty years later Maltby \shortcite{Maltby71} discovered that the magnetic splitting of Li\,{\sc i} 6708\AA\, resonance doublet in sunspots shows a Paschen-Back pattern.

In the analysis and modelling of spectral lines observed in magnetic upper main sequence stars, it is normally assumed that the splitting of the lines is correctly described by the anomalous Zeeman effect. However, in some cases the fine structure splitting of one or both levels involved in a transition is very small, and magnetic fields found in some stars are large enough to produce magnetic splitting comparable in size to this small fine structure splitting. In these cases, the usual anomalous Zeeman effect for isolated energy levels no longer describes the splitting correctly. Instead, the splitting of the line should be calculated taking into account both the fine structure splitting and the magnetic splitting simultaneously. This regime is known as the {\it incomplete} or {\it partial  Paschen-Back} (PPB) effect. The partial Paschen-Back regime can occur when observed spectral line profiles are created by closely spaced doublets or triplets. A number of spectral lines which are important for analysis of magnetic upper main sequence stars have such structure, for example the triplet lines of He~{\sc i}, the 6707\AA\, resonance lines of Li~{\sc i}, the 4481\AA\, line of Mg~{\sc ii}, and numerous lines such as $\lambda\lambda$~4128, 4200, 4621, 5466 and 7849 of Si~{\sc ii}. In the presence of a sufficiently strong magnetic field (typically some kG), the magnetic splitting for such lines can become comparable to or larger than the value of the fine structure splitting, and the resulting pattern for the split components can differ significantly from the one calculated according to the usual anomalous Zeeman effect.

Thus, it is natural to suppose that examples of the partial Paschen-Back pattern could be found in the stellar spectra of stars with strong magnetic fields. Most such stars belong to the class of chemically peculiar (CP or Ap-Bp) stars, and for them it is important to correctly model magnetic line splitting in order to obtain realistic results from studies of magnetic field structure and from abundance analysis.
%Polosukhina Li\,{\sc i} $\lambda$6708\AA, Mathys Fe\,{\sc ii} $\lambda$6147\AA\,, $\lambda$6149\AA\,
The first study of how the Paschen-Back splitting affects the Stokes profiles of spectral lines formed under the typical conditions of magnetic Ap and Bp stars was carried out by Mathys (1990) in his study of the line profiles of the Fe~{\sc ii} multiplet 74 doublet $\lambda\lambda$~6147-49 in stars with fields of several kG. A general discussion of the theoretical methods used to describe the transition from the Zeeman to the Paschen-Back effect in conditions appropriate to such stars was presented by Landolfi et al. \shortcite{Landolfi+01}.

During the last decade, several articles have been published on the simulation of line profiles taking into account the partial Paschen-Back effect. Such work includes analysis of the Li\,{\sc i} 6708\AA\, resonance doublet \cite{Kochukhov08}, of the He\,{\sc i} 10830\AA\, multiplet \cite{Socas-Navarro+04}, and %hyperfine structure (),
the components of the Fe\,{\sc ii} multiplet 74 (Stift et al. 2008).
%\cite{Mathys90}, \cite{Stift+08}.
The influence of the Paschen-Back effect on the scattering polarization of molecular lines was considered by Shapiro et al. \shortcite{Shapiro+06}.

Because of the importance of silicon in the spectrum of magnetic CP stars, and the fact that many of the commonly analysed spectral lines of Si~{\sc ii} are closely spaced fine structure doublets, we have carried out a study of the PPB magnetic splitting and of the resulting
Stokes $I$ and $V$ profiles for a number of strong and weak Si\,{\sc ii} and Si\,{\sc iii} lines that are observed in the spectra of magnetic CP stars in the spectral band between 3800\AA\, and 8000\AA. Typically, such stars show enhanced silicon abundance in stellar atmospheres (Landstreet et al. 1989), %(Renson \& Manfroid 2009; Paunzen et al. 2011),
and the location of their overabundance spots depends on the configuration of the magnetic field structure (Hatzes 1991, 1997; Khochukhov 2002; Shavrina et al. 2010; L\"uftinger et al. 2010).
%correlate (L\"uftinger) or anticorrelate (Shavrina)?
Therefore, these profiles are modified by the presence of a complex magnetic field structure and a non-uniform abundance distribution of silicon.
Practical attempts to simultaneously fit several silicon lines in the spectrum of the magnetic CP star HD~318107, which has a field of order 15~kG,  show that, assuming Zeeman splitting, some lines can be fit well in the framework of a chosen model for magnetic field structure, while the other lines, such as Si\,{\sc ii} 5041\AA\, and 5056\AA, can not \cite{Bailey+11}. This is an example of a modelling problem which suggests that the theoretical description of line splitting in a magnetic field could introduce unnecessary ambiguity into modelling. In an effort to understand to what extent the difference between Zeeman splitting theory and that of the PPB theory is an important source of modelling errors, we have carried out a series of calculations to see how the Stokes $I$ and $V$ profiles differ when these lines are computed according to the partial Paschen-Back effect or the Zeeman effect.

A description of the method for evaluating the difference between simulated profiles assuming the partial Paschen-Back effect and the Zeeman effect is given in Sec.~\ref{method}, and the results of its application to a number of Si\,{\sc ii} and Si\,{\sc iii} lines are presented in Sec.~\ref{results}. Section~\ref{special} is devoted to analysis of the characteristics of the Si\,{\sc ii} 5041\AA\, line magnetically widened in the regime of the partial Paschen-Back effect. The best fit results for the Stokes $IV$ profiles of this line in HD~318107 spectra are also presented in that section for simulations that employ the partial Paschen-Back effect and the Zeeman effect. Discussion of the results obtained is given in Sec.~\ref{summary}.

\section{Method}
\label{method}

The theory behind the calculation of partial Paschen-Back splitting can be found in Chapter 3 of the recent monograph on polarization in spectral lines by Landi Degl'Innocenti and Landolfi \shortcite{LDL04}. The energy levels are calculated by diagonalization of the total Hamiltonian, which consists of unperturbed and magnetic parts, under the assumption that the atomic system is described by the L-S coupling scheme. For the purposes of this paper, the unperturbed energy levels (for zero magnetic field) of different ions are taken from the NIST atomic database \cite{Ralchenko+11} and the oscillator strengths of the respective transitions are taken from the NIST and VALD \cite{Kupka+00} atomic databases.

%Introduction of the partial Paschen-Back effect.
For weak magnetic fields the results of the Zeeman and Paschen-Back splitting computations are almost the same. With an increase of the field strength the magnetic splitting becomes comparable to or even larger than the fine-structure separation between J-levels (where J stands for the total angular momentum of the atomic system). This condition gives rise to the partial Paschen-Back effect, which results in wavelengths and intensities of the split components which differ significantly from the predictions of the Zeeman effect for the same strength and direction of the applied magnetic field. Furthermore, in the PPB regime transitions with $|\Delta J| > 1$ are possible. In strong magnetic fields such forbidden transitions produce spectral lines of relatively strong intensity. Such forbidden multiplet components can be observed in the spectra of some Ap stars \cite{Mathys90} and have been called  ``ghost" lines (Landi Degl'Innocenti \& Landolfi 2004; Stift et al. 2008).

In this work the numerical calculation of Paschen-Back splitting is carried out using equation (3.64) from the aforementioned monograph. The code (by V.K.) has been designed for calculation of the Paschen-Back splitting and has been verified through comparison of the results obtained with numerous examples (splitting of the terms $^2$P, $^2$D, $^2$F, $^3$P, $^3$D)\footnote{
The splitting of the higher terms (up to $^8$H) has also been calculated to verify the code and the similarity of splitting patterns.} from Landi Degl'Innocenti and Landolfi \shortcite{LDL04}, with the results of splitting of the Li\,{\sc i} 6708\AA\, resonance doublet \cite{Kochukhov08} and with previous computations of the components of the Fe\,{\sc ii} multiplet 74 \cite{Stift+08}.
%For example, the upper panel of Fig.~\ref{fig1a} shows a comparison between the Paschen-Back and Zeeman magnetic splitting of the lowermost two levels of the $^4$D term for the Fe\,{\sc ii} multiplet 74. There is no splitting of the energy level around 31368.5 cm$^{-1}$ with J=$\frac{1}{2}$ in the case of the Zeeman effect due to a zero Lande factor, while the Paschen-Back effect results in the magnetic splitting of this level and is similar to that obtained by Stift et al. \shortcite{Stift+08}. The Paschen-Back splitting of the so called ``ghost" component at $\lambda$6408.9\AA\, of the Fe\,{\sc ii} multiplet 74 as a function of the magnetic field strength is shown in the lower panel of Fig.~\ref{fig1a}. These results are similar to those obtained in the work of Stift et al. \shortcite{Stift+08}.
The splitting and relative intensities of lines from Table~\ref{tab1} in the regime of the partial Paschen-Back effect are consistent with the results obtained by using the code designed by Landi Degl'Innocenti (private communication) %\shortcite{Landi11}
for calculation of Paschen-Back splitting. Thus, our method gives results similar to those previously published.

\subsection{Calculation of Paschen-Back splitting}
\label{procedure}

After the verification of the part of the code designed to calculate Paschen-Back splitting, it was incorporated into the ZEEMAN2 code. This code was created by Landstreet \shortcite{Landstreet88} for the simulation of polarimetric (Stokes $IVQU$ ) line profiles, and was later modified by Khalack \& Wade \shortcite{Khalack+Wade06}, who added an automatic minimization of the model parameters using the downhill simplex method \cite{press+}.

\begin{table*}
\parbox[t]{\textwidth}{
\centering
\caption[]{The air wavelengths and oscillator strengths of Si\,{\sc ii} and Si\,{\sc iii} lines analyzed, assuming no magnetic field.}
\begin{tabular}{cccrrrccccrcccc}
\hline
\hline
Multiplet&$\lambda$, \AA& \multicolumn{3}{c}{$\log~gf$}&\multicolumn{5}{c}{Lower term (NIST)}  &\multicolumn{5}{c}{Upper term (NIST)}\\ \cline{2-14}
%\multicolumn{2}{c}{Energy, cm$^{-1}$ (NIST)}\\ \cline{2-6}
number&   & L-S & NIST & VALD & Energy,& S & L & J& Land\'{e} & Energy,& S & L & J& Land\'{e} \\
      &   &coupling & & & cm$^{-1}$& & & &factor& cm$^{-1}$& & & &factor \\
\hline
\multicolumn{14}{c}{Si\,{\sc ii}} \\
1.00&3853.6645 & -1.3887& -1.341& -1.517& 55309.35& 0.5& 2.0& 1.5&0.800& 81251.32& 0.5& 1.0& 1.5& 1.177 \\
    &3856.0176 & -0.4342& -0.406& -0.557& 55325.18& 0.5& 2.0& 2.5&1.200& 81251.32& 0.5& 1.0& 1.5& 1.177 \\
    &3862.5954 & -0.6895& -0.757& -0.817& 55309.35& 0.5& 2.0& 1.5&0.800& 81191.34& 0.5& 1.0& 0.5& 0.333 \\
    &3864.9593 & -9.9999& & & 55325.18& 0.5& 2.0& 2.5& & 81191.34& 0.5& 1.0& 0.5&  \\
\\
3.00&4128.0536 &  0.3813&  0.359&  0.316& 79338.50& 0.5& 2.0& 1.5&0.800& 103556.16& 0.5& 3.0& 2.5& 0.857 \\
    &4128.0758 & -9.9999& & & 79338.50& 0.5& 2.0& 1.5& & 103556.03& 0.5& 3.0& 3.5& \\
    &4130.8715 & -0.7648& -0.783& -0.824& 79355.02& 0.5& 2.0& 2.5&1.200& 103556.16& 0.5& 3.0& 2.5& 0.857 \\
    &4130.8937 &  0.5362&  0.552&  0.476& 79355.02& 0.5& 2.0& 2.5&1.200& 103556.03& 0.5& 3.0& 3.5& 1.143 \\
\\
7.06&4200.6578 & -0.8883& -0.889& -0.820& 101023.05& 0.5& 2.0& 1.5&0.800& 124822.14& 0.5& 3.0& 2.5& 0.857 \\
    &4200.6684 & -9.9999& & & 101023.05& 0.5& 2.0& 1.5& & 124822.08& 0.5& 3.0& 3.5& \\
    &4200.8873 & -2.0344& -2.034& -1.970& 101024.35& 0.5& 2.0& 2.5&1.200& 124822.14& 0.5& 3.0& 2.5& 0.857 \\
    &4200.8979 & -0.7334& -0.733& -0.670& 101024.35& 0.5& 2.0& 2.5&1.200& 124822.08& 0.5& 3.0& 3.5& 1.143 \\
\\
7.05&4621.4181 & -0.6079& -0.608& -0.540& 101023.05& 0.5& 2.0& 1.5&0.800& 122655.37& 0.5& 3.0& 2.5& 0.857 \\
    &4621.4439 & -9.9999& & & 101023.05& 0.5& 2.0& 1.5&0.800& 122655.25& 0.5& 3.0& 3.5& \\
    &4621.6960 & -1.7540& -1.754& -1.680& 101024.35& 0.5& 2.0& 2.5&1.200& 122655.37& 0.5& 3.0& 2.5& 0.857 \\
    &4621.7216 & -0.4530& -0.453& -0.380& 101024.35& 0.5& 2.0& 2.5&1.200& 122655.25& 0.5& 3.0& 3.5& 1.143 \\
\\
    &5040.6934 & -9.9999& & & 81191.34& 0.5& 1.0& 0.5& & 101024.35& 0.5& 2.0& 2.5& \\
5.00&5041.0238 &  0.3229&  0.029&  0.291& 81191.34& 0.5& 1.0& 0.5&0.333& 101023.05& 0.5& 2.0& 1.5&0.800 \\
    &5055.9842 &  0.5781&  0.523&  0.593& 81251.32& 0.5& 1.0& 1.5&1.177& 101024.35& 0.5& 2.0& 2.5&1.200 \\
    &5056.3166 & -0.3761& -0.492& -0.359& 81251.32& 0.5& 1.0& 1.5&1.177& 101023.05& 0.5& 2.0& 1.5&0.800\\
\\
7.03&5466.4607 & -0.2369& -0.237& -0.200& 101023.05& 0.5& 2.0& 1.5&0.800& 119311.34& 0.5& 3.0& 2.5& 0.857 \\
    &5466.5055 & -9.9999& & & 101023.05& 0.5& 2.0& 1.5& & 119311.19& 0.5& 3.0& 3.5& \\
    &5466.8493 & -1.3830& -1.383& -1.340& 101024.35& 0.5& 2.0& 2.5&1.200& 119311.34& 0.5& 3.0& 2.5& 0.857 \\
    &5466.8942 & -0.0820& -0.082& -0.040& 101024.35& 0.5& 2.0& 2.5&1.200& 119311.19& 0.5& 3.0& 3.5& 1.143 \\
\\
4.00&5957.5588 & -0.2983& -0.225& -0.301& 81191.34& 0.5& 1.0& 0.5&0.333& 97972.09& 0.5& 0.0& 0.5&2.000 \\
    &5978.9297 &  0.0027&  0.084&  0.004& 81251.32& 0.5& 1.0& 1.5&1.177& 97972.09& 0.5& 0.0& 0.5&2.000 \\
\\
2.00&6347.1088 &  0.2973&  0.149&  0.297& 65500.47& 0.5& 0.0& 0.5&2.000& 81251.32& 0.5& 1.0& 1.5&1.177 \\
    &6371.3715 & -0.0037& -0.082& -0.003& 65500.47& 0.5& 0.0& 0.5&2.000& 81191.34& 0.5& 1.0& 0.5&0.333 \\
\\
7.02&7848.8165 &  0.3155&  0.316&  0.330& 101023.05& 0.5& 2.0& 1.5&0.800& 113760.32& 0.5& 3.0& 2.5& 0.857 \\
    &7848.9213 & -9.9999& & & 101023.05& 0.5& 2.0& 1.5& & 113760.15& 0.5& 3.0& 3.5& \\
    &7849.6177 & -0.8307& -0.831& -0.810& 101024.35& 0.5& 2.0& 2.5&1.200& 113760.32& 0.5& 3.0& 2.5& 0.857 \\
    &7849.7224 &  0.4704&  0.470&  0.490& 101024.35& 0.5& 2.0& 2.5&1.200& 113760.15& 0.5& 3.0& 3.5& 1.143 \\
\hline
\multicolumn{14}{c}{Si\,{\sc iii}} \\
2.00&4552.6216 & 0.2911&  0.292& 0.181& 153377.05& 1.0& 0.0& 1.0& 1.000& 175336.26& 1.0& 1.0& 2.0& 1.500 \\
    &4567.8403 & 0.0692&  0.068& -0.039& 153377.05& 1.0& 0.0& 1.0& 1.000& 175263.10& 1.0& 1.0& 1.0& 1.500 \\
    &4574.7570 & -0.4079& -0.409& -0.509& 153377.05& 1.0& 0.0& 1.0& 1.000& 175230.01& 1.0& 1.0& 0.0& 1.500 \\
\\
8.09&4716.6540 &  0.4910&  0.491& 0.440& 204330.79& 0.0& 2.0& 2.0& 1.000& 225526.33& 0.0& 3.0& 3.0& 1.000 \\
\\
4.00&5739.7300 &  0.2920&  0.292& -0.160& 159069.61& 0.0& 0.0& 0.0& 1.500& 176487.19& 0.0& 1.0& 1.0& 1.000 \\
\hline
\label{tab1}\end{tabular}
}
\end{table*}

To find the partial Paschen-Back splitting for even a small range of wavelengths, all the components of the multiplet under consideration must be taken into account because the intensity normalization of the split components is done over the whole multiplet. The procedure for calculation of Paschen-Back splitting takes into account the unperturbed (or magnetically perturbed) energy levels and determines the respective air wavelength and oscillator strength of components, based on the term configurations and the total strength of all lines in the multiplet under consideration.
With this code it is possible to simulate a line profile composed of several spectral lines (blends), some of which are split in the Paschen-Back regime, while the others are split in the Zeeman regime. One of the two routines (Paschen-Back or Zeeman) is applied to each of the contributions to the blends.

To find the difference between the spectral profiles simulated assuming the PPB and Zeeman regimes, it is essential to require that both sets of computations are carried out with identical atomic data. Under the assumption of zero magnetic field there should be no difference between these profiles. The procedure for calculating partial Paschen-Back splitting recalculates the wavelengths and intensities of all the line components for each field strength, including zero field. Therefore,  for a correct comparison of the two cases it is necessary to adopt the relative line strengths within the multiplet, as derived in the PPB case for zero magnetic field, for the calculation of Zeeman splitting of the same set of spectral lines. For the condition of zero magnetic field there is no actual splitting in the PPB regime, but the relative intensity of each line in a multiplet is determined through the sum of its components assuming the Russell-Saunders (or L-S) coupling scheme for the energy levels \cite{LDL04}. This same set of zero-field relative intensities must be adopted for the Zeeman splitting case. However, the total intensity of the multiplet is normalised to the sum of oscillator strengths of its members taken from the NIST or, sometimes, VALD atomic databases. The  adopted oscillator strength of each line is determined by multiplication of its relative strength in the multiplet by this sum (see the third column in Table~\ref{tab1}).

The oscillator strengths derived from L-S coupling for spectral lines in the multiplets studied here generally show good agreement with the NIST data. However, the VALD data have been used for overall multiplet normalisation when they provided better agreement with the L-S coupling relative intensities of spectral lines in the multiplet studied. In Table~\ref{tab1} the oscillator strengths from VALD are employed for simulation of Si\,{\sc ii} lines $\lambda\lambda$~5041, 5056, 5957, 5979, 6347 and 6371, and the remaining oscillator strengths are derived from the NIST database. Note that the perturbed part of the analyzed Hamiltonian includes only the contribution from the magnetic field and is nil when the field vanishes. All other possible contributions to the perturbed part of the Hamiltonian are not considered here. Due to the simplified approach of the description of the analyzed Hamiltonian, some lines ($\lambda\lambda$~3856, 3862, 4128, etc.) still show small differences between the values tabulated in NIST or VALD and the computed values of oscillator strength.

To ensure that the ``ghost" lines play a negligible role in the case of zero magnetic field, their oscillator strengths are set to have $\log gf$ =-9.999 when the field is zero for the purposes of simulations. It should be pointed out that not all multiplets have ``ghost" lines in the framework of the Paschen-Back splitting theory, because some multiplets do not involve any pairs of energy levels with $|\Delta J| \geq 2$.

\subsection{Comparison of the Paschen-Back and Zeeman profiles}
\label{comparison}

The spectral lines studied are listed in Table~\ref{tab1}. This table collects (separately for Si~{\sc ii} and Si~{\sc iii}) the multiplet number according to Adelman et al.~\shortcite{Adelman+85},
the air wavelengths of the lines, the $\log gf$ values obtained from the NIST and VALD databases (columns 4 and 5) and the adopted oscillator strengths from L-S coupling theory, normalised to the best matching database values, in column 3. The remaining columns list the basic atomic parameters of the lower and upper energy levels involved in each transition, and the Land\'{e} splitting factors used for Zeeman splitting computations.

\begin{figure*}
\includegraphics[scale=0.82,angle=-90]{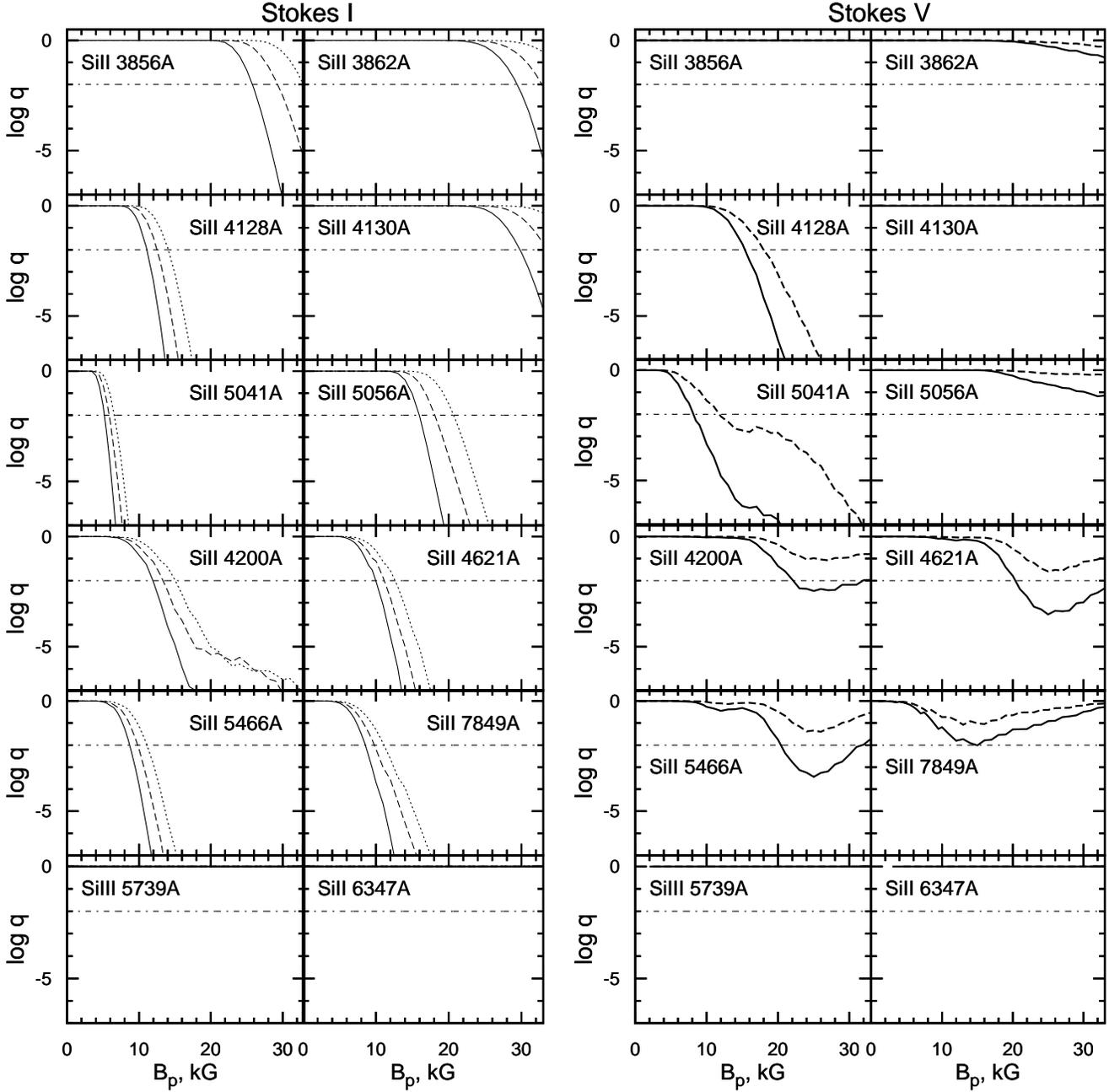}
\caption{ Logarithm of the probability that the noise with $\sigma=0.04$ (S/N=250) is masking the difference between the Stokes $I$ and $V$ profiles calculated assuming PPB and Zeeman splitting, assuming $\log{[N_{\rm Si}/N_{\rm H}]}=-3.5$ and $V \sin{i}$ = 1 km s$^{-1}$. The continuous, dashed, and dotted lines correspond to the cases where the axis
of magnetic dipole forms an angle $\alpha$= 0$^{\circ}$, 45$^{\circ}$, and 90$^{\circ}$ with the line of sight, respectively.
The horizontal dash-dotted line corresponds to the probability $q$=0.01.
The bottom panels present the examples of Si\,{\sc ii} and Si\,{\sc iii} lines that show $q\approx1$ for both Stokes $I$ and $V$ profiles with little or no dependence on the field strength.
}
\label{fig1}
\end{figure*}

The simulation of Stokes $I$ and $V$ profiles is carried out for the lines in Table~\ref{tab1}, assuming a star with effective temperature $T_{\rm eff}$=13000K, gravity $\log{g}$=4.0, zero microturbulent velocity, and an oblique rotator model with a dipolar magnetic field structure. To explore the dependence of the final results on the field orientation, the calculations have been repeated for three cases where the axis of the magnetic dipole forms an angle $\alpha$= 0$^{\circ}$, 45$^{\circ}$ and 90$^{\circ}$ with the line of sight, and the results are presented by the continuous, dashed and dotted lines, respectively (see Fig.~\ref{fig1}). To investigate the influence of the axial stellar rotation and other non-magnetic phenomena that contribute to the widening of a line profile, the analysis is also performed for different values of $V\sin{i}$ (see Fig.~\ref{fig3}). Because most of the magnetic CP stars have strong Si\,{\sc ii} lines due to an enhanced silicon abundance (see, for example, Landstreet et al. 1989),
%Renson \& Manfroid 2009; Paunzen et al. 2011),
the value $\log{[N_{\rm Si}/N_{\rm H}]}=-3.5$ is adopted here for the simulations. The silicon abundance is assumed to be homogeneously distributed in the stellar atmosphere.

The observed Stokes $I$ and $V$ profiles are usually contaminated by observational noise. For practical reasons, it is useful to compare the difference between the simulated PPB and Zeeman profiles to the noise level $\sigma \approx \frac{1}{S/N}$ at the analyzed spectral line. To decide whether this difference can be confidently detected above the noise level we use the {\it chi square probability function} and calculate the $\chi^2$ test statistics given by:

\begin{equation}\label{eq2}
%d = \sqrt{\frac{1}{n(n-1)} \sum^n_{i=1}  \frac{(I^{PB}_i - I^{Ze}_i)^2}{\sigma^2}}
\chi^2 = \frac{1}{n\sigma^2} \sum^n_{i=1}  (I^{PB}_i - I^{Ze}_i)^2 \;,
\end{equation}

\noindent where $I^{\rm PB}_i$ and  $I^{\rm Ze}_i$ represent the intensity of the Stokes $I$ (or $VQU$) profiles at a wavelength point $i$ calculated with the assumption of the PPB and Zeeman splitting, respectively. The sum is evaluated over the wavelength range within which the intensities of both profiles are significantly different from the continuum (or from zero in the case of Stokes $VQU$ profiles), which means that $|I^{\rm PB}_i|>\sigma$ and/or $|I^{\rm Ze}_i|>\sigma$.
For a particular line this test quantifies how much its PPB profile deviates from its Zeeman profile distorted by the noise. If this distortion reaches the level of difference between the PPB and Zeeman profiles the noise will mask it.
The probability $q$ that the simulated PPB profile is indistinguishable from the simulated Zeeman profiles plus noise with the dispersion $\sigma$ can be found from the chi-square probability function $p(\chi^2|\nu)$ \cite{Abramowitz+Stegun72} as:

\begin{equation}\label{eq3}
 q(\chi^2|\nu) = 1-p(\chi^2|\nu) = Q\left(\frac{\nu}{2},\frac{\chi^2}{2}\right) \;,
\end{equation}

\noindent where $\nu$ is the number of resolved elements in the analyzed profile and $Q$ denotes the {\it upper incomplete gamma function}:

\begin{equation}\label{eq4}
Q\left(\frac{\nu}{2},\frac{\chi^2}{2}\right) = \left[\Gamma\left(\frac{\nu}{2}\right)\right]^{-1} \int_{\chi^2/2}^{\infty}{t^{\frac{\nu}{2}-1} e^{-t}\; dt}
\end{equation}

\begin{figure*}
\includegraphics[scale=0.82,angle=-90]{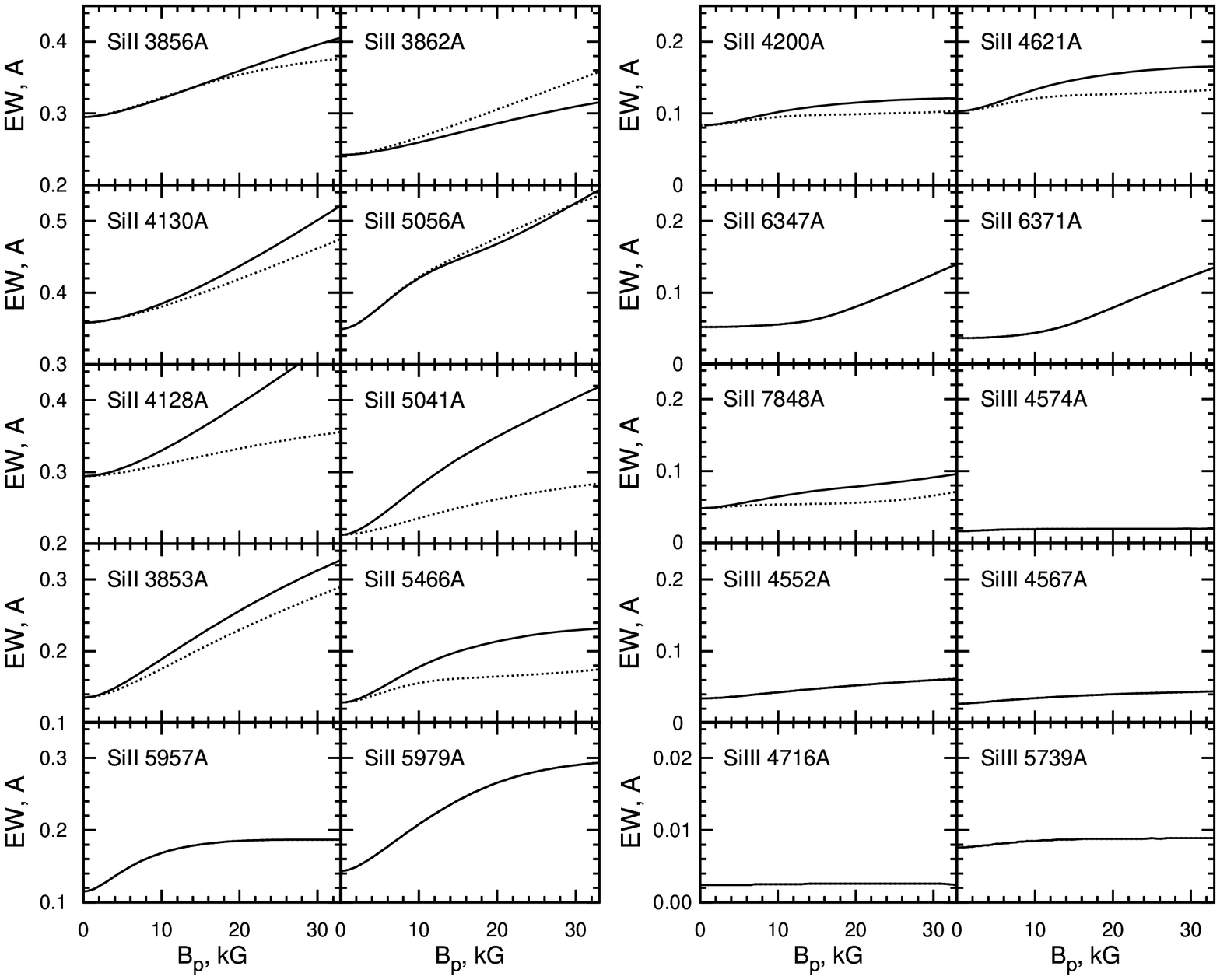}
\caption{ The equivalent widths of the analyzed lines as a function of magnetic field, assuming  a silicon abundance of $\log{[N_{\rm Si}/N_{\rm H}]}=-3.5$. The continuous and dotted lines correspond to the profiles calculated using PPB and Zeeman splitting respectively. For Si\,{\sc iii} lines 4716\AA\, and 5739\AA\, note that the vertical scale is ten times smaller. }
\label{fig2}
\end{figure*}

For a small magnetic field the $\chi^2$ value is small and the probability $q$ that the noise masks the difference between the PPB and Zeeman profiles is high. Nevertheless, $\chi^2$ grows with increased field strength and for $\chi^2=35.7$ (assuming $\nu=20$) the aforementioned probability drops below 0.01. This confidence level seems to be quite robust to perform an evaluation of the difference between the PPB and Zeeman profiles for individual lines (see, for instance, Donati et al.~\shortcite{Donati+92}). If this criterion is satisfied ($q\leq0.01$), the choice of magnetic splitting regime (PPB or Zeeman) during the simulation of the spectral line under consideration influences the results of spectral analysis.

Since the ESPaDOnS spectropolarimeter is one of the most popular instruments designed to perform spectral observations of Stokes $IVQU$ parameters, its spectral resolution $R= 65000$ and signal to noise ratio in polarimetric mode\footnote{
For more details about this instrument, visit
{\rm http://www.cfht.hawaii.edu/Instruments/Spectroscopy/Espadons/}} are adopted here for our analysis.
The given spectral resolution provides approximately 20 resolved elements in the Stokes $I$ profile of the analyzed lines. This number increases with the magnetic field strength as the Stokes $I$ profile becomes wider.

Depending on the purpose of the spectropolarimetric observations, the S/N ratio in the Stokes $I$ spectra routinely obtained with the ESPaDOnS, NARVAL and MuSiCoS (the prototype of ESPaDOnS) is between 100 and 1000 (Folsom et al. 2010; Wade et al. 2000a). Using the ESPaDOnS spectropolarimeter some observers even work with S/N$\simeq$1300 (or more) in the Stokes $I$ spectra in order to detect weak magnetic fields in massive stars \cite{Grunhut+12} or in order to reach a higher precision in reconstruction of a magnetic field configuration and/or abundance maps. % (see Sec.~\ref{special}). }
To provide the same approach to the investigation of different spectral lines, an average value S/N=250 ($\sigma$=0.004) in the Stokes $I$ spectrum is adopted here for all spectral regions.

\section{Results of analysis}
\label{results}

The list of Si\,{\sc ii} and Si\,{\sc iii} lines that are strongest in Stokes $I$ spectra is given in Table~\ref{tab1} for a star with $T_{\rm eff}$=13000K and $\log{g}$=4.0. The same list of lines is also used to analyze the Stokes $V$ line profiles.

A magnetically widened profile sometimes blends together all lines of a multiplet (for example Si\,{\sc ii} 5466\AA, 7849\AA, etc.). This demands their simultaneous analysis. In other cases the Stokes $I$ and $V$ profiles are analyzed separately for each line of the multiplet. In this study, the weak lines and the ``ghost" lines are analyzed only if they contribute to a profile formed by a strong spectral line.

\subsection{General results for a slowly rotating star}
\label{slow}

The difference between simulated profiles assuming the PPB and Zeeman splitting is first analyzed for a slowly rotating star assuming $V\sin{i}$= 1 km s$^{-1}$ and a silicon abundance of $\log{[N_{\rm Si}/N_{\rm H}]}=-3.5$. The oblique rotator model with a dipolar magnetic field is applied to describe the configuration of a magnetic field. The parameter $\alpha$ defines the angle between the dipole magnetic axis and the line of sight. The field strength at the poles of the magnetic dipole considered here changes gradually from zero up to 30 kG (for $B_{\rm p}\leq$ 10 kG the step in the field strength is 500 G and for $B_{\rm p}>$ 10 kG it is 1 kG). Note that a polar field strength $B_{\rm p}$ of 10~kG corresponds typically to an observed mean field modulus $\langle |B| \rangle$ of 7 or 8~kG, and to a mean longitudinal field strength $\langle B_z \rangle$ of about 3~kG or less, so this polar field strength is typical of quite a number of known magnetic ApBp stars.

Depending on the sensitivity of the probability logarithm (see Eq.~\ref{eq3}) to the field strength at the magnetic poles, all the analyzed lines can be divided into two groups. The first group includes spectral lines for which the probability $q$ calculated for the Stokes $I$ %and $V$
profiles decreases strongly with the magnetic field strength (see Fig.~\ref{fig1}). Some of these lines are comparatively strong (Si\,{\sc ii} $\lambda\lambda$~3853, 3856, 3862, 4128, 4130, 5041, 5056) with a relative intensity in Stokes $I$ spectra of less than 0.6 (for $V\sin{i}$= 1 km s$^{-1}$), while the others are weak (Si\,{\sc ii} $\lambda\lambda$~4200, 4621, 5466, 7849). Except for the Si\,{\sc ii} lines $\lambda\lambda$~3853, 3856, 3862, 4130 and 5056 all of them show an average difference between the Paschen-Back and Zeeman profiles {that can be masked by the noise with the probability $q<0.01$ (assuming S/N=250) in the Stokes $I$ spectrum for a magnetic field strength $B_{\rm p}$=5 - 15 kG.}
%that exceeds the level of significance (for S/N=250) in the Stokes $I$ spectrum
For the Si\,{\sc ii} lines $\lambda\lambda$~3856, 3862 and 4130 there is no dependence of $\log q$ on magnetic field strength for $B_{\rm p}<$20 kG (see the left panel of Fig.~\ref{fig1}).

The slope of the $\log q$ with respect to the strength of magnetic field is smaller for Stokes $V$ profiles than for Stokes $I$ profiles for these lines. Assuming the S/N=250 for the data in Stokes $V$ spectra, the  Si\,{\sc ii} lines $\lambda\lambda$~4128 and 5041 show an average difference between the Paschen-Back and Zeeman profiles that exceeds the level of significance for $B_{\rm p}>$15 kG and $B_{\rm p}>$7 kG, respectively (see the right panel of Fig.~\ref{fig1}). In the case of Si\,{\sc ii} line 5041\AA\, the difference appears mainly due to the presence of a comparatively strong ``ghost" line in the PPB splitting (see Sec.~\ref{special}).
For the Stokes $VQU$ spectra the S/N ratio averaged over the available line profiles is usually much less than the value obtained for the Stokes $I$ spectrum \cite{Donati+97}. According to the definition of the Stokes parameters (see for details chapter 1 in Landi Degl'Innocenti and Landolfi \shortcite{LDL04}) the Stokes $VQU$ spectra are obtained as a difference of two measurements taken with different combinations of orientation of a quarter-wave retarder and/or a polarizer \cite{Donati+99}. In this case the resulting signal is significantly smaller comparing to the Stokes $I$ spectrum, while the average uncertainty remains the same. Therefore, if we adopt S/N=50 ($\sigma$=0.02) for the Stokes $VQU$ spectra there will be no significant difference between the simulated Paschen-Back and Zeeman profiles (the probability $q$ is close to 1).

For weak magnetic fields there is no significant difference in the behaviour of the $\log q$ for different angles $\alpha$. %between the dipole magnetic axis and the line of sight.
However, for $B_{\rm p}>$~ 5~kG the $\log q$ decreases faster with the magnetic field strength for $\alpha$= 0$^{\circ}$ than for $\alpha$= 45$^{\circ}$ for both Stokes $I$ and $V$ profiles. The main reason is that the contribution to line profiles from the area of visible stellar disc that is close to the equator of magnetic dipole (where the field is weaker than near the poles), grows with the angle $\alpha$ producing higher $\chi^2$ (see Eq.~\ref{eq2}). For the case when $\alpha$= 90$^{\circ}$, there is no data for the Stokes $V$ profiles (see the right panel of Fig.~\ref{fig1}), because the signal becomes too weak (smaller than $\sigma$) for this orientation of the magnetic dipole. For the Stokes $I$ profiles and $\alpha$= 90$^{\circ}$ one can see that the decrease of the $\log q$ with the magnetic field strength is less steep compared the other two cases.

\begin{figure}
\includegraphics[scale=0.68,angle=-90]{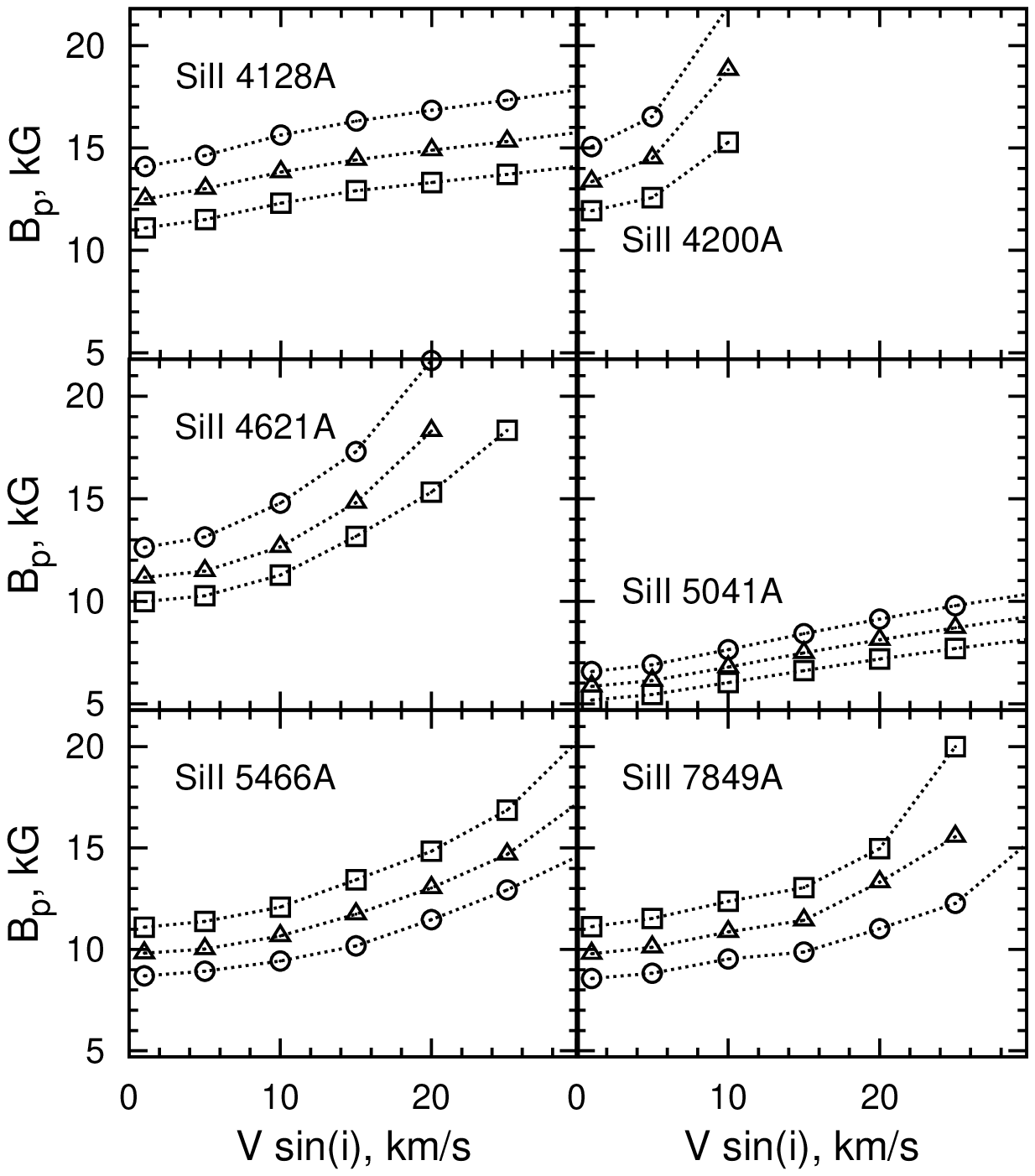}
\caption{Variation of the threshold of the magnetic field (see details in Subsec.~\ref{rapid}) with $V\sin{i}$ for the Stokes $I$ spectra.
%(left panel) and $V$ (right panel) spectra.
The open circles, triangles, and squares correspond to the cases where the axis of magnetic dipole forms an angle $\alpha$= 0$^{\circ}$, 45$^{\circ}$, and 90$^{\circ}$ with the line of sight, respectively. }
\label{fig3}
\end{figure}

The dependence of the equivalent widths (EW) on the magnetic field strength for most of the lines listed in Table~\ref{tab1} is shown in Fig.~\ref{fig2}. In general, for lines from the first group, the equivalent widths of Stokes $I$ profiles calculated employing PPB splitting grow faster with the magnetic field (continuous line) than the equivalent widths of profiles calculated assuming Zeeman splitting (dotted line). The Si\,{\sc ii} lines $\lambda\lambda$~4128, 5041 and 5466 show a difference between the equivalent widths of the PPB and Zeeman profiles $\Delta EW >$ 0.020~\AA\ for a magnetic field $B_{\rm p}\geq 10$~kG. This fact indicates that when the magnetic field at the magnetic poles reaches a strength of 10~kG, the difference between the PPB and Zeeman profiles in unpolarized spectra for the aforementioned lines is quite significant (see also Fig. ~\ref{fig7}).

The spectral lines for which the $\log q$ shows very weak or no dependence on the magnetic field strength for the Stokes $I$ and $V$ profiles (see the bottom panels of Fig.~\ref{fig1}) compose the second group. One reason for the occurrence of such spectral lines is that for transitions between terms with spin quantum number S=0, the pattern of the PPB splitting is the same as the one obtained in the Zeeman splitting \cite{LDL04}.
The Si\,{\sc iii} lines 5739\AA\, and 4716\AA\, (see Tabl.~\ref{tab1}), for which $\chi^2=0$ for all values of the magnetic field strength (see Eq.~\ref{eq2} and the corresponding panels for Si\,{\sc iii} line 5739\AA\ in Fig.~\ref{fig1}) are good examples of this rule.

The other lines (Si\,{\sc ii} $\lambda\lambda$~5957, 5979, 6347, 6371 and Si\,{\sc iii} $\lambda\lambda$~4552, 4567, 4574) provide $\log q$ close to zero with almost no dependence relative to the magnetic field strength for the Stokes $I$ and $V$ profiles (see panels for Si\,{\sc ii} line 6347\AA\, in Fig.~\ref{fig1}). These lines belong to widely spaced multiplets, and so even for field strengths $B_{\rm p}\simeq 30$~kG they are still well in the Zeeman regime. We see clearly that spectral lines belonging to the second group can be analyzed without any loss of information assuming Zeeman splitting even for stars with very strong magnetic fields.

The Stokes $Q$ and $U$ profiles are quite sensitive to the configuration of the magnetic field and to the abundance inhomogeneities \cite{Wade+00a}. Nevertheless, with the assumption of a S/N=250 for Stokes $Q$ and $U$ spectra and a homogeneous distribution of silicon abundance, the Stokes $Q$ and $U$ profiles appear to be rather weak.
The analyzed silicon lines do not show any significant differences between the Stokes $Q$ and $U$ profiles computed with the Zeeman effect or the PPB effect for $B_{\rm p}<$ 30 kG and they can be simulated assuming the Zeeman splitting without any loss of information. With the assumption of a S/N$<$50 the Stokes $Q$ and $U$ profiles are barely detectable, hence
in further analysis, most attention is focussed only on the Stokes $I$ and $V$ profiles computed according to the Zeeman and PPB effects.

\begin{figure*}
\includegraphics[scale=0.34,angle=-90]{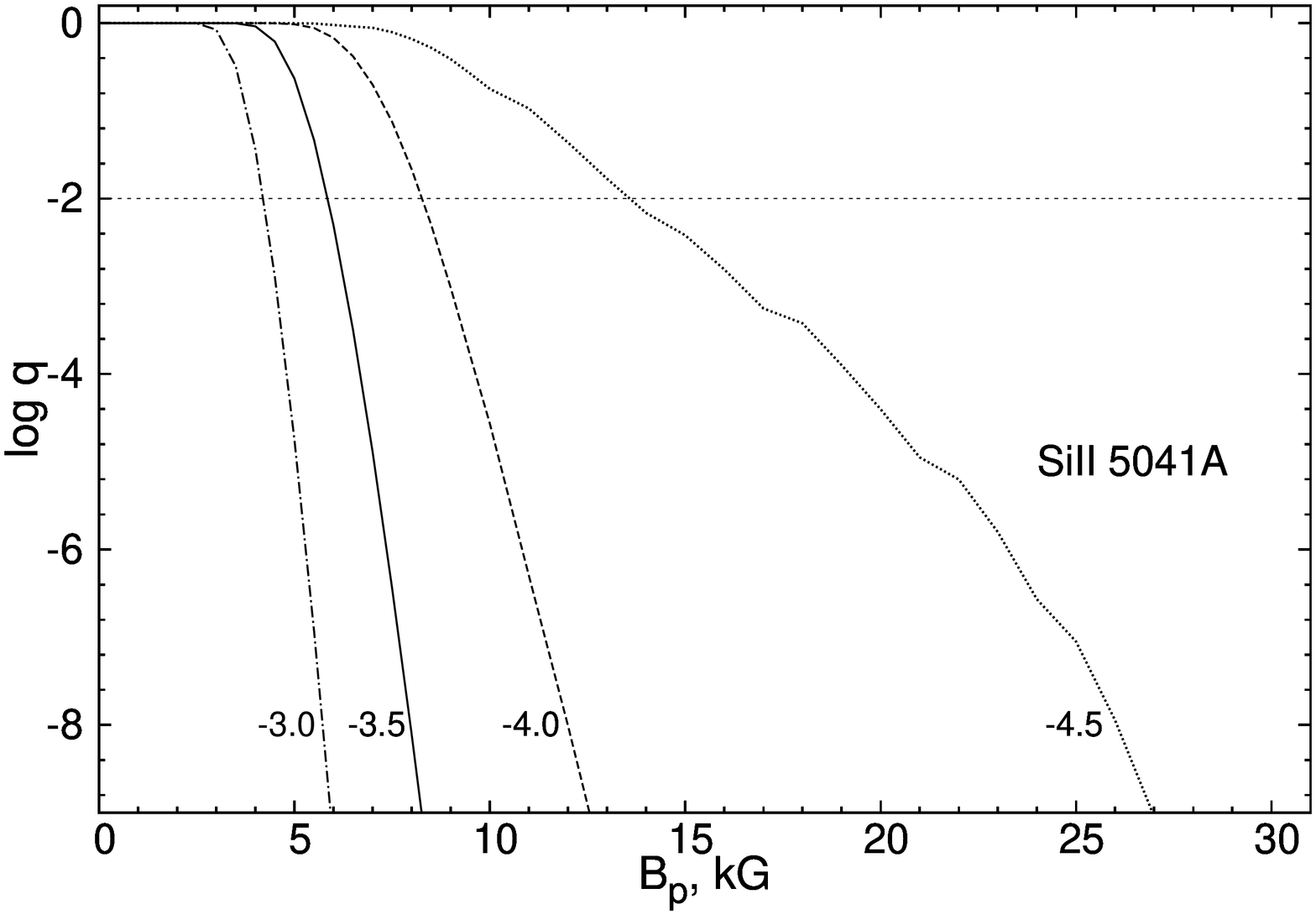}
\includegraphics[scale=0.34,angle=-90]{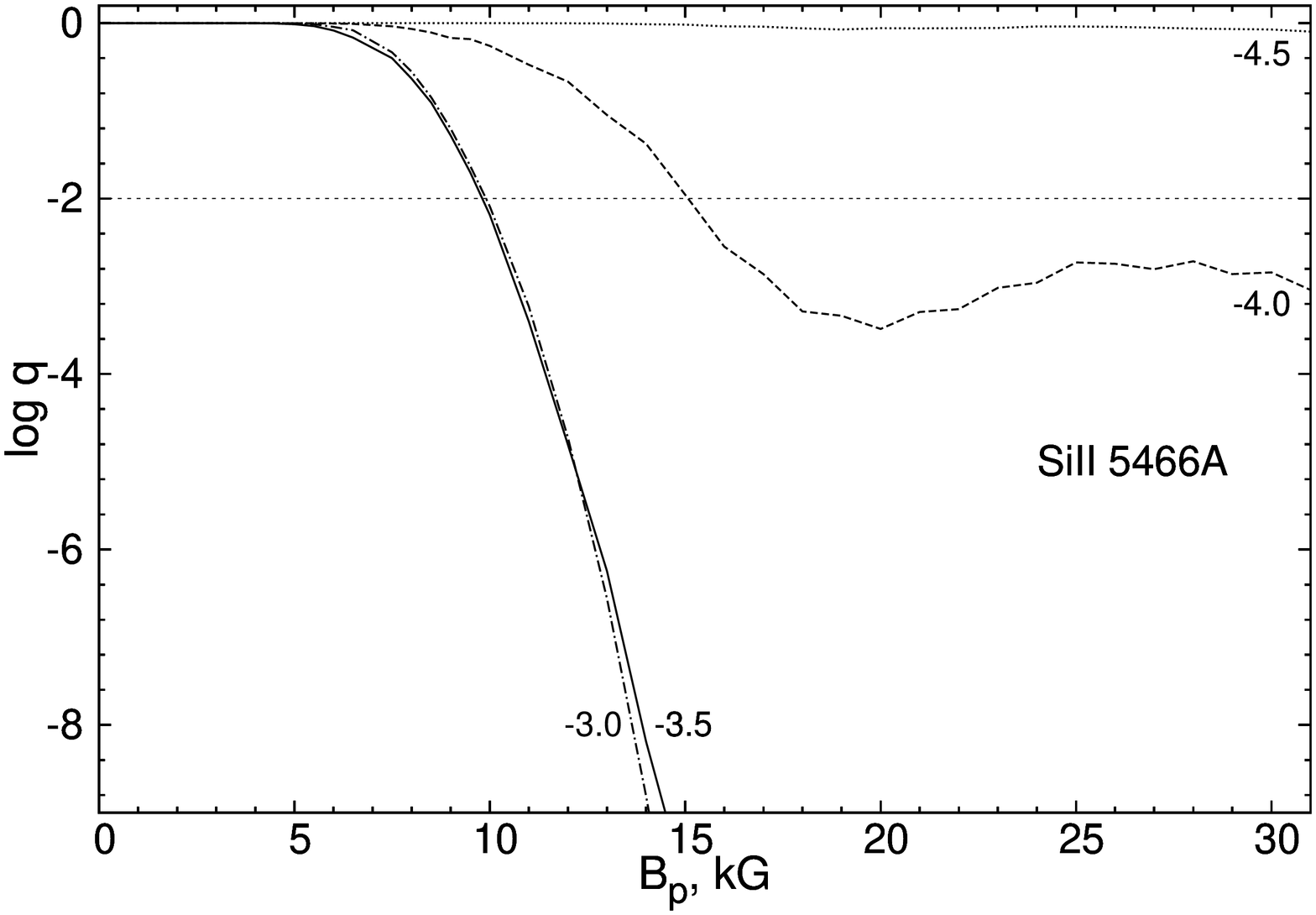}
\caption{Difference between the Stokes $I$ profiles calculated with the PPB and Zeeman splitting
for the Si\,{\sc ii} lines 5041\AA\, (left panel) and 5466\AA\,
(right panel) for different silicon abundances (see the values of $\log{[N_{\rm Si}/N_{\rm H}]}$ on the panels),
assuming a S/N=250 and an angle $\alpha$= 45$^{\circ}$ between the line of sight and the axis of the magnetic dipole.
The dashed horizontal lines present the level of significance for p=0.99 (q=0.01, see Eq.~\ref{eq3}).
 }
\label{fig4}
\end{figure*}

\subsection{Dependence on $V\sin{i}$}
\label{rapid}

The majority of magnetic CP stars show velocities of axial rotation with $V\sin{i}$ greater than 1 km s$^{-1}$, and mostly less than 50 km s$^{-1}$ (Kudryavtsev et al. 2007; Wade et al. 2000b).
The dependence of the magnetic field strength at which $\log q=-2$ on the projected rotation velocity is studied here for some lines in order to evaluate its influence on the $\chi^2$ (see Eq.~\ref{eq2}). The strength of the magnetic field determined in this way provides the threshold above which the line profile simulated in the PPB regime becomes significantly different from the one simulated assuming Zeeman splitting. % for a particular spectral line.

The values of the polar magnetic field corresponding to $\log q=-2$ are shown in Fig.~\ref{fig3} for several lines assuming S/N=250. For the Si\,{\sc ii} lines  $\lambda\lambda$~4128 and 5041, the derived threshold grows slowly with $V\sin{i}$ for the Stokes $I$ spectra and corresponds to field strength around 10-18 kG and around 5-10 kG, respectively.
For Si\,{\sc ii} lines $\lambda\lambda$~4621, 5466 and 7849 the derived threshold increases as a function of $V\sin{i}$, but remains less than 18 kG for $V\sin{i}<$ 15 km s$^{-1}$. Meanwhile, for Si\,{\sc ii} 4200\AA\, the derived threshold increases rapidly as a function of $V\sin{i}$.
Similar results are derived for two other values of the angle $\alpha$ (see Fig.~\ref{fig3}).

The results obtained suggest that it may be important to take into account the PPB effect during the simulation of the Si\,{\sc ii} lines of the first group, especially the line 5041\AA, for magnetic stars with $V\sin{i}<$~15 km s$^{-1}$ and $B_{\rm p}>$ 10-18 kG.

\subsection{Dependence on silicon abundance}
\label{abun}

Some magnetic CP stars show patches of an enhanced silicon abundance in their stellar atmospheres (Hatzes 1997; Khochukhov 2002; Shavrina et al. 2010; L\"{u}ftinger et al. 2010; Bailey et al. 2011). An increase of silicon abundance leads to stronger line profiles when keeping the other parameters constant. Two examples of how the dependence of $\log q$ on the magnetic field strength changes with increasing silicon abundance (homogeneously spread over stellar surface) are shown in Fig.~\ref{fig4} for the Si\,{\sc ii} lines 5041\AA\, (left panel) and 5466\AA\, (right panel), belonging to the first group
(see Subsec.~\ref{slow}). For clarity only the curves that correspond to the model with an angle $\alpha=$ 45$^{\circ}$ between the line of sight and the axis of the magnetic dipole are shown, while the values of adopted silicon abundance $\log{[N_{\rm Si}/N_{\rm H}]}$ are placed on the panels near each curve. Similar behaviour of the $\log q$ is found for other values of the angle $\alpha$.

In the case of Si\,{\sc ii} 5041\AA\, line, an increase of silicon abundance leads to a steeper dependence of the logarithm of probability $\log q$ on the magnetic field strength and correspondingly to a smaller magnetic field at which $\log q=-2$ and the difference between the PPB and Zeeman profiles becomes significant. This behaviour
can be explained as the effect of intensification of the ``ghost" line (see Sec.~\ref{special} and Fig.~\ref{fig7}) with increasing silicon abundance, which in turn results in a bigger difference between the PPB and Zeeman profiles (larger $\chi^2$ in Eq.~\ref{eq2}). A similar behaviour can also be seen in the case of the weak line Si\,{\sc ii} 5466\AA. Nevertheless, for a silicon abundance $\log{[N_{\rm Si}/N_{\rm H}]}= -4.0$ the dependence of the $\log q$ on the magnetic field strength changes its slope for $B_{\rm p}>$ 20 kG, while for $\log{[N_{\rm Si}/N_{\rm H}]}= -4.5$ the logarithm of probability $\log q \simeq 0$ for $B_{\rm p}\leq$ 30 kG.

The first example illustrates the unique behaviour of the Si\,{\sc ii} 5041\AA\, line, while the second example shows a typical example of how a dependence of $\log q$ on the magnetic field strength changes with increased silicon abundance. % depending on the relative intensity and position of split components in the PPB and Zeeman regimes.
The results obtained here indicate that variation of silicon abundance by about 1~dex does not eliminate the difference between the PPB and Zeeman profiles; in general, this difference grows with increased silicon abundance.

\section{The Si\,{\sc ii} line 5041\AA}
\label{special}

For most of the Si\,{\sc ii} and Si\,{\sc iii} spectral line profiles studied, here a contribution from their respective ``ghost" components remains relatively small when the magnetic field strength is $B_{\rm p}<$ 30 kG. The same result was found by Stift et al. \shortcite{Stift+08} during the analysis of PPB splitting of the Fe\,{\sc ii} multiplet 74.

\begin{figure}
\includegraphics[scale=0.35,angle=-90]{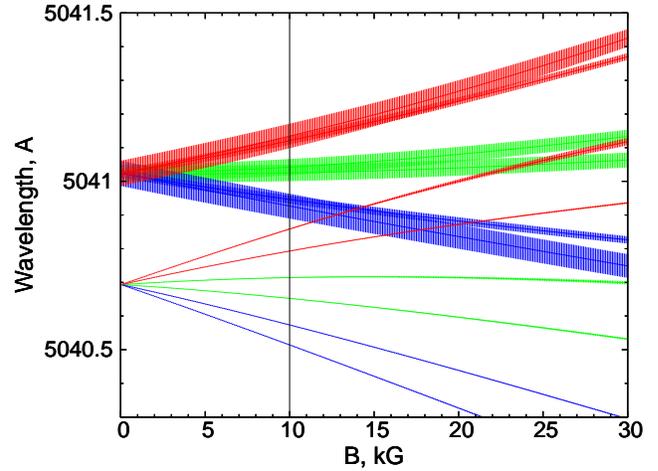}
\caption{(Colour online) Splitting and relative intensities of the subcomponents of the Si\,{\sc ii} lines 5041.0\AA\, and 5040.7\AA\, (``ghost" line) in the PPB regime. The magnetic field strength of 10~kG is represented by the vertical line and the $\sigma_{\rm blue}$, $\pi$, and $\sigma_{\rm red}$ components are plotted respectively in blue, green, and red.}
\label{fig5}
\end{figure}

However, this is not true for the Si\,{\sc ii} line 5041\AA. From the left panel of Fig.~\ref{fig4} we can see that the PPB effect becomes important for simulations of the Stokes $I$ profile even for solar silicon abundance when the field strength at the pole exceeds 14 kG. Fig.~\ref{fig5} shows the wavelength splitting and relative intensities of the subcomponents of the Si\,{\sc ii} line 5041.0\AA\, and the ``ghost" line at 5040.7\AA. It appears that at $B_{\rm p}\simeq$ 12 kG those components start overlapping, while the $\pi$ and $\sigma_{\rm red}$-components of the ``ghost" line increase in intensity rapidly with the strength of the magnetic field.

For a smaller magnetic field ($B_{\rm p}<$ 12 kG) the main difference between the Paschen-Back and Zeeman profiles comes from the depression caused by the ``ghost" line. As an example, Fig.~\ref{fig7} shows the comparison of Paschen-Back and Zeeman profiles in the Stokes $I$ (left panel) and $V$ (right panel) spectra simulated assuming enhanced silicon abundance $\log{[N_{\rm Si}/N_{\rm H}]}=$ -3.5 with $B_{\rm p}$= 10 kG and $\alpha$=45$^{\circ}$ for most of the analyzed lines. In the case of Si\,{\sc ii} 5041\AA\, line the ``ghost" component is located around $V$= - 20 km s$^{-1}$ and its contribution to the left wing of the line is significant. It increases with the silicon abundance (see left panel of Fig.~\ref{fig4}) and can be incorrectly identified as a blend if one uses the Zeeman effect approximation to simulate the Stokes $I$ and $V$ profiles of this particular line. A spectral analysis of the depression caused by the ``ghost" component is a good tool for estimating the silicon abundance and determining the magnetic field configuration of stars with spectra clearly showing the presence of silicon lines.

\subsection{The 5041\AA\, line in the spectrum of HD~318107}
\label{star1}
To test the theoretical results obtained for the Si\,{\sc ii} line 5041\AA\, the Stokes $I$ and $V$ profiles of this particular line and of the line Si\,{\sc ii} 6347\AA\, were analyzed for the magnetic Ap star HD318107 using both Zeeman and PPB splitting. The 6347\AA\ line is chosen as a comparison, because its Zeeman pattern of split components does not differ much from the pattern of PPB splitting for any magnetic field strength $B_{\rm p}<$ 30 kG, and correspondingly the parameter $d$ is close to zero (see Sec.~\ref{slow}). The polarimetric spectra were obtained at the rotational phase 0.991 using the echelle spectrapolarimeter ESPaDOnS at CFHT with a resolving power of R $\simeq$ 65000 and signal-to-noise ratio S/N $\simeq$ 200
% for which area?
\cite{Bailey+11}.

\begin{figure*}
\includegraphics[scale=0.34,angle=-90]{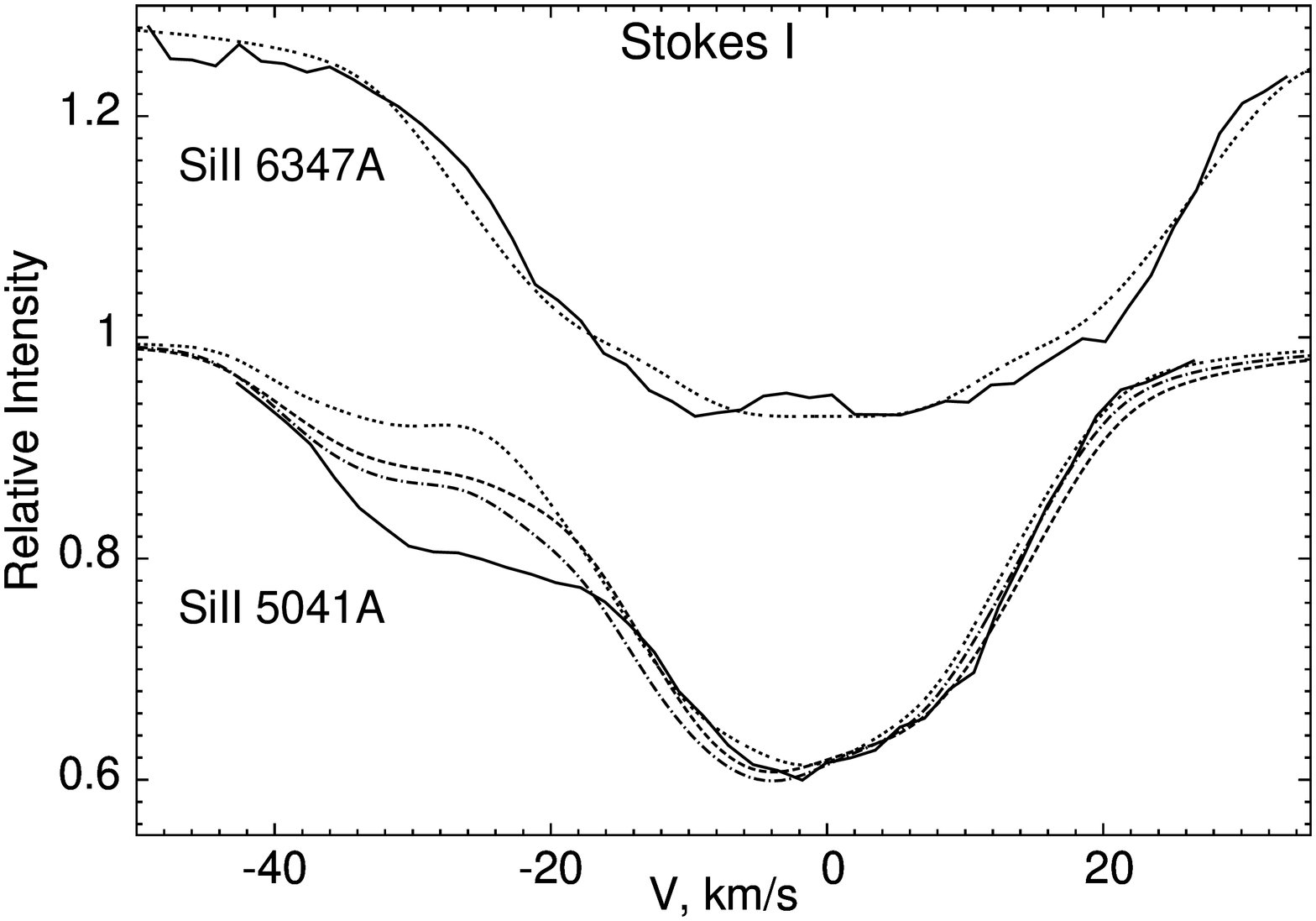}
\includegraphics[scale=0.34,angle=-90]{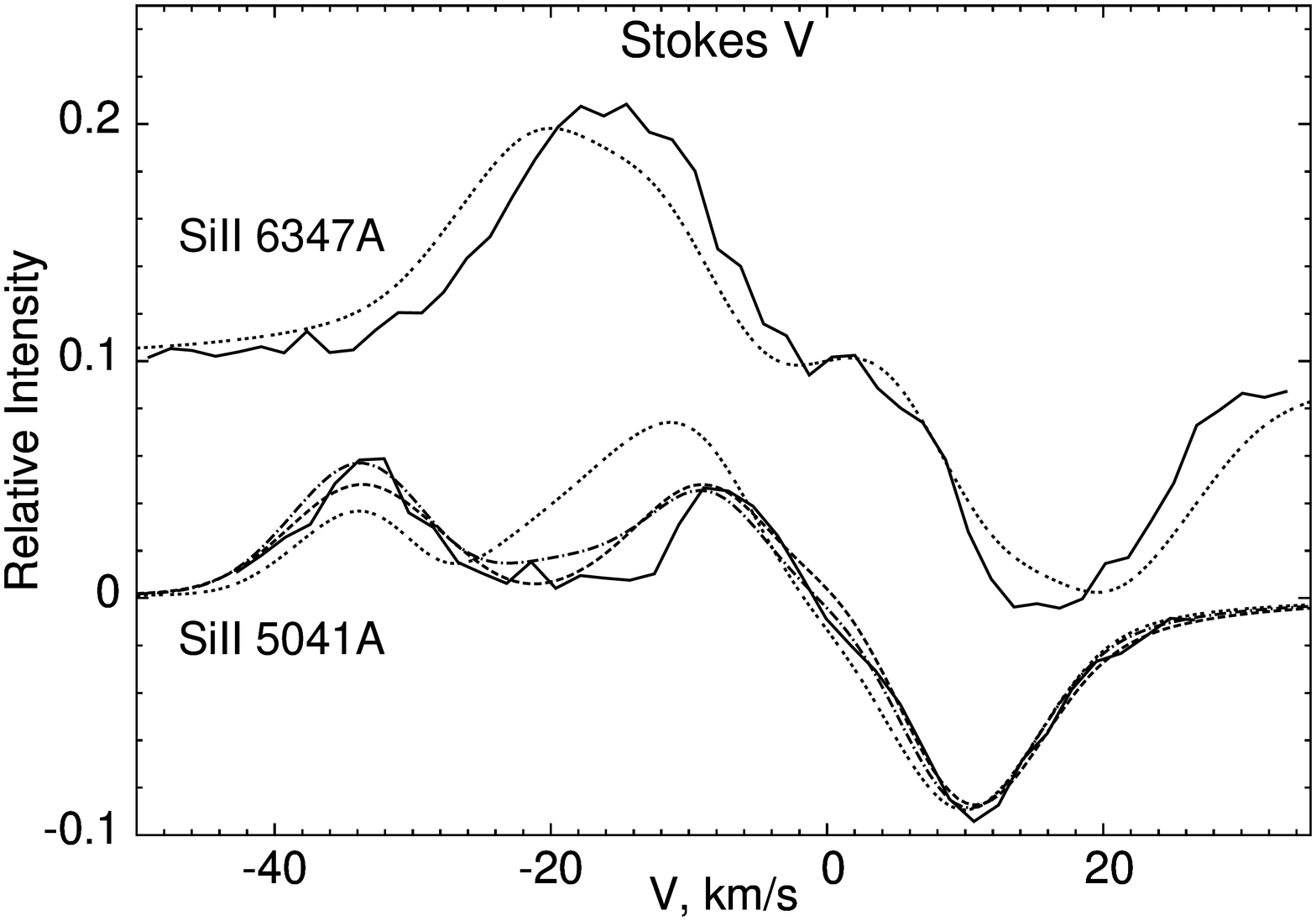}
\caption{The best fit model for the Stokes $I$ (left) and Stokes $V$ (right) profiles of Si\,{\sc ii} 5041\AA\, and 6347\AA\, lines assuming Zeeman (dotted line) and PPB (dashed line) splitting. Continuous lines represent the observed data, while the dash-dotted lines provides the best fit results for Si\,{\sc ii} 5041\AA\, in the PPB regime in combination with the Fe\,{\sc i} and Fe\,{\sc ii} blends in the Zeeman regime for $\log{[N_{\rm Fe}/N_{\rm H}]}=-3.0$ . In the case of Si\,{\sc ii} 5041\AA\, the dotted line shows the best fit results for the silicon and iron lines split in the Zeeman regime. The results for the Zeeman splitting of this line without contribution from the iron lines are not shown here. The 6347\AA\, spectra are shifted vertically for better visibility.}
\label{fig6}
\end{figure*}

HD318107 is a very peculiar magnetic star with a rotational period of $P = 9.7088 \pm 0.0007$ days, deduced by
%Manfroid \& Mathys~\cite{M+M00}
Bailey et al. \shortcite{Bailey+11} from analysis of the periodic variability of its magnetic field components. For the rotational phase analysed here its mean longitudinal magnetic field $\langle B_z \rangle$ is around 5 kG and the mean field modulus $\langle |B| \rangle$ is about 14 kG. This star has an effective temperature $T_{\rm eff}$= 11800$\pm$500 K and gravity $\log{g}$= 4.22$\pm$0.13 \cite{Landstreet+07}. The stellar atmosphere model was calculated for temperature $T_{\rm eff}$= 11800 K and gravity $\log{g}$= 4.2 using the {\sc Phoenix} code \cite{Hauschildt+97} assuming LTE and solar metallicity. The modified ZEEMAN2 code (see Sec.~\ref{procedure}) was employed to analyze the spectropolarimetric data assuming a homogeneous distribution of silicon abundance and a magnetic field geometry characterized by a sum of colinear dipole, quadrupole, and octupole at an angle $\beta$ to the rotational axis. To describe the geometry of a magnetic field we have adopted from Bailey et al. \shortcite{Bailey+11} the following parameters : $i$=22\degr, $\beta$=65\degr, $B_{\rm d}$= 25.6 kG, $B_{\rm q}$= -12.8 kG, $B_{\rm o}$= 0.9 kG.

Results of the best fit assuming Zeeman magnetic splitting for both lines and PPB splitting for the Si\,{\sc ii} line 5041\AA\, are shown in the Table~\ref{tab2}, where the chi-squared values are given for combined analysis of the Stokes $I$ and $V$ profiles.
For each case listed in this table the fitting of silicon abundance, $V \sin(i)$ and radial velocity has been performed individually using the downhill simplex method \cite{press+}.
In the case of the Si\,{\sc ii} 6347\AA\, line, application of the Zeeman splitting approximation results in a good fit (see Fig.~\ref{fig6}), while in the case of Si\,{\sc ii} 5041\AA\, line, PPB splitting provides a much better fit than can be obtained with a Zeeman profile.

The weak blending lines Fe\,{\sc i} 5040.85\AA, 5040.90\AA\, and Fe\,{\sc ii} 5040.76\AA\, also can make a contribution to the observed profile in the area of the ``ghost" component of Si\,{\sc ii} 5041\AA\, line. To analyze their influence on the best fit results, these lines were included in the computations, and split in the Zeeman regime during the simulation routine, while the Si\,{\sc ii} line 5041\AA\, was split in the PPB regime. The respective best fit results are given at the second line of Table~\ref{tab2} (see dash-dotted line at the Fig.~\ref{fig6}) and are much better in the PPB regime than the results obtained for the same set of lines split in the Zeeman regime (dotted line).

\begin{table}
%\parbox[t]{\textwidth}{
\centering
\caption[]{Approximation of Si\,{\sc ii} 5041\AA\, and 6347\AA\, lines observed in the spectrum of HD~318107 (phase 0.991) employing the PPB and Zeeman effects for their magnetic splitting and adopting from Bailey et al. \shortcite{Bailey+11} %$\log[N_{\rm Si}/N_{\rm H}]=-3.65$, $V \sin(i)$ = 7 km s$^{-1}$, RV = -9 km s$^{-1}$, and
the following parameters for the magnetic field geometry: $i$=22\degr, $\beta$=65\degr, $B_{\rm d}$= 25.6 kG, $B_{\rm q}$= -12.8 kG, $B_{\rm o}$= 0.9 kG.}
\begin{tabular}{cccccr}
\hline
\hline
$\lambda$, & Splitting& $\log[N_{\rm Si}/N_{\rm H}]$ & $V \sin(i)$, & Vr, & $\chi^2/\nu$ \\
\AA & & & km s$^{-1}$& km s$^{-1}$& \\
\hline
%multicolumn{11}{c}{HD~318107} \\
5041& PPB    & -3.57$\pm$0.15& 7.9$\pm$1.0 & -8.4$\pm$1.0 & 16.5  \\
5041& PPB+Fe & -3.80$\pm$0.15& 7.3$\pm$1.5 & -8.1$\pm$1.0 & 9.8   \\
5041& Zeeman & -3.74$\pm$0.15& 7.5$\pm$0.8 & -8.0$\pm$1.0 & 70.7  \\
5041& Zeeman+Fe & -3.94$\pm$0.15& 7.5$\pm$0.8 & -7.7$\pm$1.0 & 40.4  \\
6347& Zeeman & -3.67$\pm$0.15& 7.4$\pm$1.0 & -8.9$\pm$1.0 & 7.8   \\

%\multicolumn{11}{c}{HD~215441} \\
%5041& PPB & -1.3887& -1.341& -1.517& 55309.35& 0.5& 2.0& 1.5&0.800& 81251.32    \\
%6347& PPB & -0.4342& -0.406& -0.557& 55325.18& 0.5& 2.0& 2.5&1.200& 81251.32   \\
\hline
\label{tab2}\end{tabular}
%}
\end{table}

\begin{figure*}
\includegraphics[scale=0.680,angle=-90]{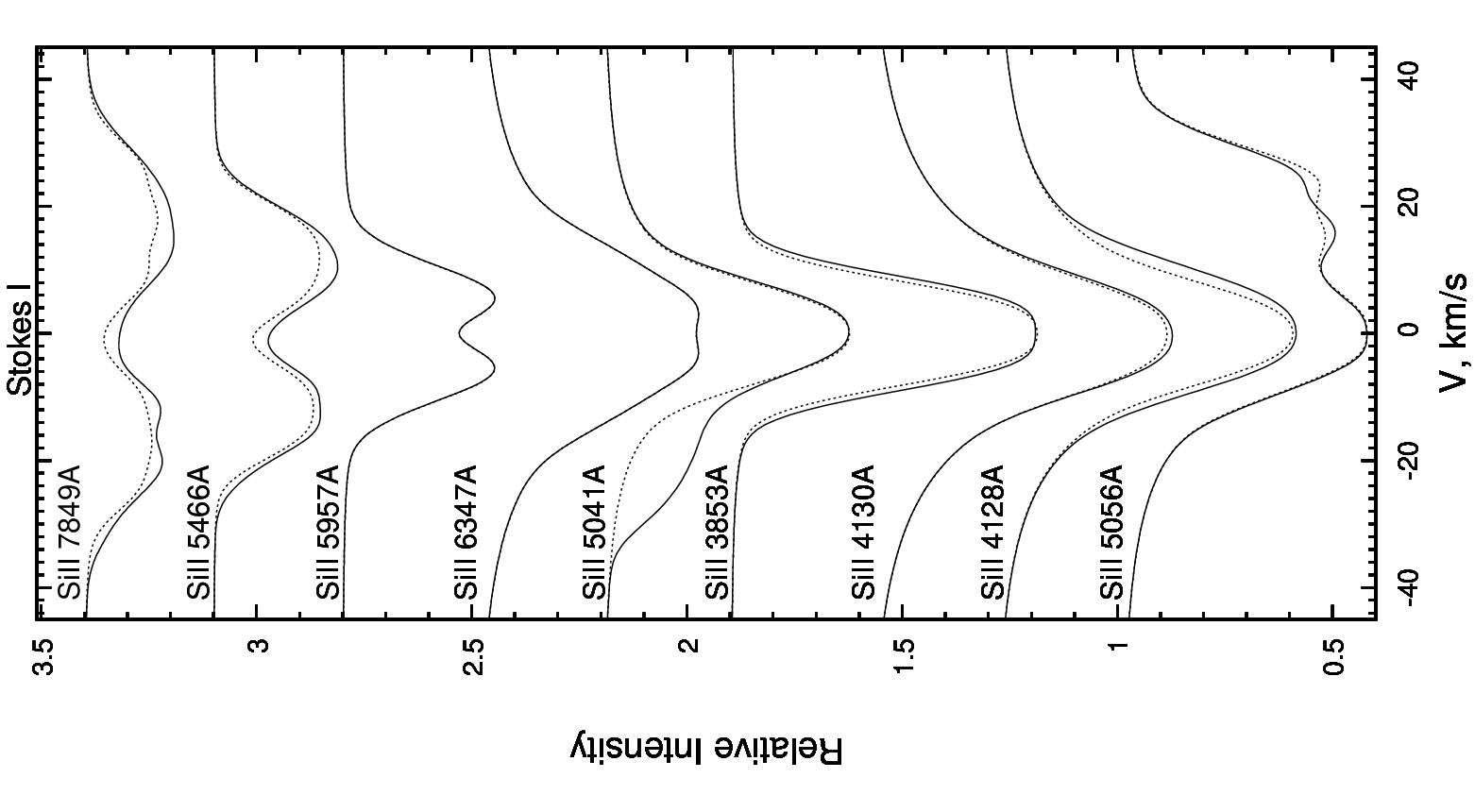}
\includegraphics[scale=0.680,angle=-90]{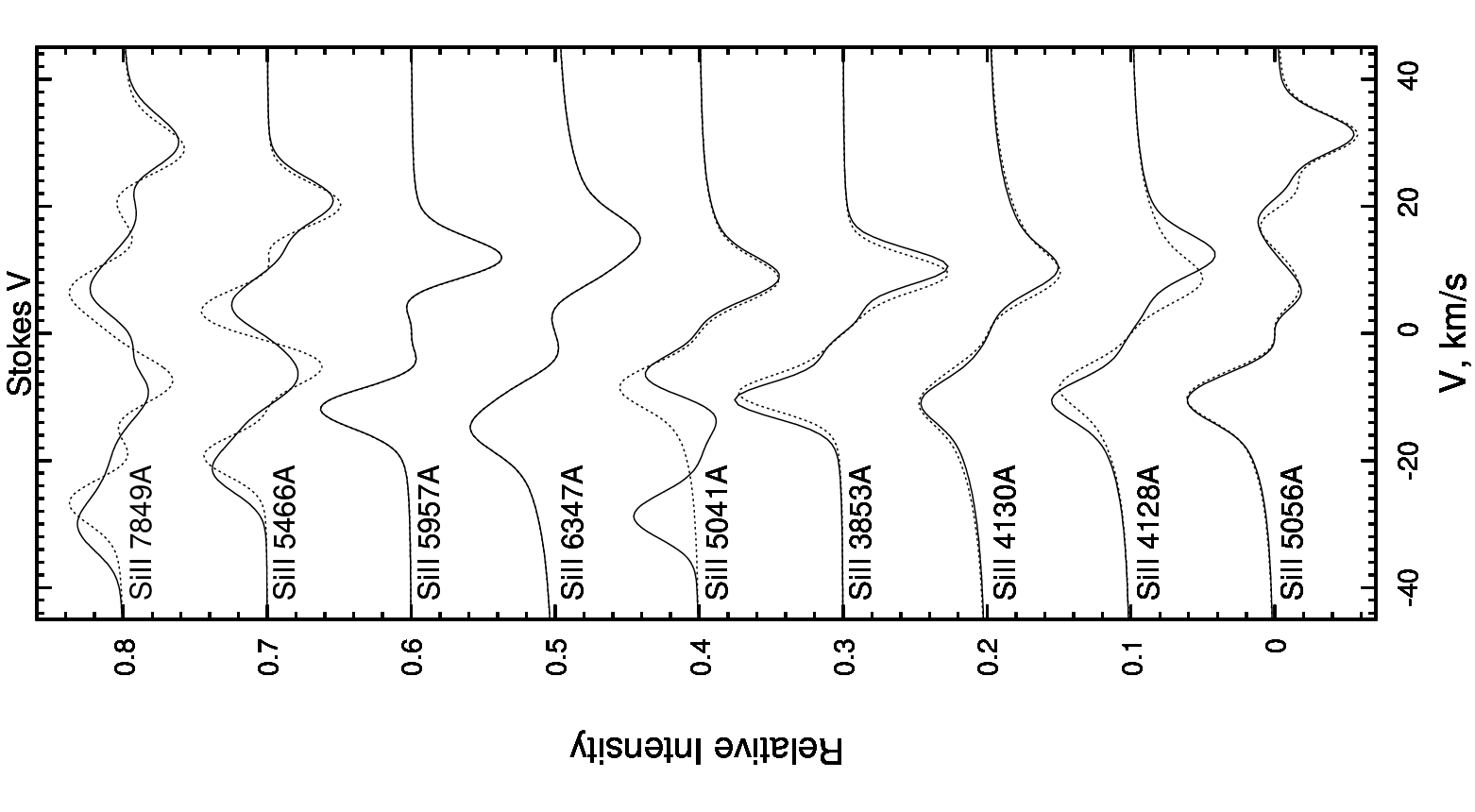}
%scale=0.34
%\special{psfile=BHB-Teff-Strat.eps angle=-90 hoffset=-10 voffset=185 vscale=33 hscale=33}
 % \hfill \\
\caption{The simulated Stokes $I$ (left panel) and $V$ (right panel) profiles of some silicon lines, applying the PPB (continuous line) and Zeeman (dotted line) splitting for an abundance $\log{[N_{\rm Si}/N_{\rm H}]}=-3.5$ and a magnetic dipole with the field strength of 10~kG at the magnetic poles. The spectra are shifted vertically for better visibility by 0.3 and 0.1 in the left and the right panels respectively. In the left wing of Si\,{\sc ii} line 5041\AA\, a contribution from the ``ghost" component is clearly visible. }
\label{fig7}
\end{figure*}

One can see that the data obtained independently for both lines agree between themselves and are close to the results derived by Bailey et al. \shortcite{Bailey+11} from a complex analysis of different spectral lines for this star.
Nevertheless, the depression in the left wing of the Si\,{\sc ii} 5041\AA\, line can not be fit sufficiently well taking into account its PPB splitting and the contribution of the iron blends. The remaining differences between the PPB profiles and the observed spectra may be partially explained in terms of more complicated magnetic field structure and/or horizontal and vertical stratification of the silicon abundance \cite{Bailey+11}.

%\section{Discussion}
%\label{discussion}

\section{Summary}
\label{summary}

This paper presents the results of simulation of Stokes $I$ and $V$ profiles of strong and weak Si\,{\sc ii} and Si\,{\sc iii} lines (see Tab.~\ref{tab1}) in the presence of stellar magnetic field. The purpose of this study is to find out the conditions under which the application of the Zeeman effect approximation is unreliable.
%in calculation of the magnetic widening of spectral lines

Presently, the Zeeman effect is used in most of the codes designed to calculate the Stokes $IVQU$  profiles of spectral lines.
For magnetic CP stars with a comparatively weak magnetic fields its application is generally valid for most of the spectral lines and results in a correct reconstruction of the abundance maps and the magnetic field configuration. Even for the stars with comparatively strong magnetic fields (20~kG~$<B_{\rm p}<$~30 kG), whose spectra have a S/N ratio around 50, the Stokes $I$  profiles of almost all the Si\,{\sc ii} lines (except 4128\AA, 5041\AA\, and 5466\AA) may be treated with the Zeeman approximation.
Nevertheless, if the available polarimetric spectra of the stars with strong magnetic field have higher S/N ratio (for example S/N=250 and more) the use of PPB splitting during the analysis of spectral lines is necessary to obtain precise results in the framework of an assumed model for the abundance map and the magnetic field structure.
In particular, Stokes $I$ %and $V$
profiles of the Si\,{\sc ii} lines which belong to the first group, as defined in Subsec.~\ref{slow}, 
%(these are Si lines for which the Zeeman splitting pattern is significantly different from that of the PPB effect), 
when calculated with the PPB splitting, differ significantly from those calculated with the Zeeman effect. This difference appears due to the different relative intensities and positions of split $\sigma$ and $\pi$-components in the PPB and Zeeman regimes, and due to the so called ``ghost" lines ($|\Delta J| \geq 2$) as in the case of Si\,{\sc ii} 5041\AA\, line (see Sec.~\ref{special}).

In our simulations (except Subsec.~\ref{star1}) we adopt a signal to noise ratio S/N= 250 in the Stokes $I$ and $V$ spectra.
% and S/N= 50 in the Stokes $VQU$ spectra.
A portion of our analysis has been performed for an enhanced silicon abundance $\log{[N_{\rm Si}/N_{\rm H}]}=-3.5$, $V\sin{i}$= 1 km s$^{-1}$.
% assuming a signal to noise ratio S/N= 500 in the Stokes {\bf $IV$ spectra}.
Increasing the rotational velocity causes a little decrease of the difference between Paschen-Back and Zeeman profiles for some Si\,{\sc ii} lines and diminishes it significantly for others. The analysis of the Si\,{\sc ii} lines $\lambda\lambda$~4128, 5041, 5466 and 7849 shows that the application of the PPB splitting for the simulation of Stokes $I$ profiles remains important in stars with $V\sin{i}<$ 15 km s$^{-1}$ when the field strength is $B_{\rm p}>$~10-18 kG.

The weak silicon lines can usually be used for spectral analysis only if the spectra are obtained with a high S/N ratio or if silicon is significantly overabundant in the stellar atmosphere.
For higher S/N ratio (S/N$>$ 250), significant differences between the Paschen-Back and Zeeman profiles can also be found for weaker magnetic fields. % (see Subsec.~\ref{comparison}).
The higher silicon abundance causes a steeper decrease of $log~q$ (the logarithm of probability that the noise masks the difference between the PPB and Zeeman profiles, see Eq.~\ref{eq3} for details) with the magnetic field strength (see Subsec.~\ref{abun}). If the patches of enhanced silicon abundance (approximately by 1 dex) provide a significant contribution to the observed line profiles, the PPB effect should be applied to correctly simulate the magnetic broadening of the Stokes $IV$ profiles of the silicon lines belonging to the first group, even for a star with a moderate magnetic field ($B_{\rm p}>$~10 kG). This statement is especially important for the rotational phases of a star with a magnetic dipole axis pointing towards the observer.

For the case of the the Si\,{\sc ii} 5041\AA\, line most of the difference between the Zeeman and Paschen-Back profiles is caused by a significant contribution from the ``ghost" line, which is well visible in the spectra with S/N=250 for $B_{\rm p}>$~6 kG and an enhanced silicon abundance (see the right panel of Fig.~\ref{fig1}). The contribution of this ``ghost" line can grow essentially with the strength of magnetic field. If one relies on the Zeeman effect during the simulation of this spectral line, the observed depression in its left wing can only be partially explained by a presence of iron blends (see Subsec.~\ref{star1}). In contrast, the use of the PPB effect approach allows for the incorporation of the ``ghost'' line and results in the much better fit of the observed in HD~318107 Stokes $IV$ profiles. A combination of the iron blend split in the Zeeman regime and the Si\,{\sc ii} 5041\AA\, line split in the PPB regime provides an even better fit (see Fig.~\ref{fig6} and Table~\ref{tab2}). Therefore, application of the PPB splitting of this particular line provides us with a good tool for estimation of the silicon abundance and for the reconstruction of the magnetic field configuration.

%The theory of the Paschen-Back effect is described in details by Landi Degl'Innocenti and Landolfi \shortcite{LDL04} for the %transitions with L-S coupling and the researcher are encouraged to use this tool for the correct simulation of Stokes $IVQU$  %profiles.

The Paschen-Back effect approximation has not been commonly used for simulations of the Stokes $IVQU$  profiles mainly because it requires more sophisticated calculations and a high precision knowledge of the energy levels for the transitions studied. Its use is becoming more common in recent astrophysical research, especially for the investigation of stellar objects with strong magnetic fields. Application of the Zeeman effect to calculate the magnetic broadening of all spectral lines in these objects can lead to imprecisions in some cases. A combination of the PPB effect analysis for the first group lines with the less demanding Zeeman effect analysis for the second group lines can provide more precise results. For stars with somewhat weaker fields, the important result of this work is to quantify the level of error introduced by using the Zeeman approximation in situations where the PPB effect should in principle be used, making it possible for investigators to decide which method is required in specific applications.

The use of this technique for the analysis of stellar spectra will be explored by the authors in the near future as well as the similar analysis of spectral lines produced by the other chemical species in which the PPB effect may play a significant role.
Meanwhile, for the silicon lines of the second group an application of the Zeeman effect provides the correct results of magnetic splitting for comparatively strong magnetic field.

\section*{Acknowledgments}

V.K. sincerely thanks  Prof. E. Landi Degl'Innocenti and Prof. F. LeBlanc for fruitful discussions and significant help in this research. J.D.L. acknowledges financial support from the Natural Sciences and Engineering Research Council of Canada. We are also grateful to the referee, who has carefully read the paper, and has made a number of very useful comments and suggestions which have improved the paper significantly.
%\vspace{0.2cm}
%\noindent
This paper has been typeset from a \TeX/\LaTeX\, file prepared by the authors.

\label{lastpage}

\end{document}